\documentclass[12pt]{article}
\usepackage{amstex}
\newcommand{\db}[1]{\bigtriangleup\beta_{{#1}}}
\newcommand{\ex}[1]{{\mathrm e}^{{#1}}}
\newcommand{\ip}[1]{\int\frac{{\mathrm d}^3{#1}}{(2\pi)^3}}
\newcommand{\ix}[1]{\int{\mathrm d}^3{#1}}
\newcommand{\ga}[2]{\int{\mathrm d}\mu_{{#1}}({#2})}
\newcommand{\kg}[1]{{\mathrm K}_{{#1}}}
\newcommand{\lo}[1]{{\mathcal L}_{#1}}
\newcommand{\ob}[2]{{\mathcal O}_{{#1}}\left({#2}\right)}
\newcommand{\po}[2]{V\left({{#1}}\vert {{#2}}\right)}
\newcommand{\rf}[1]{(\ref{#1})}
\newcommand{\rg}[1]{{\mathrm R}_{{#1}}}
\newcommand{\se}[1]{{\mathbb{#1}}}
\newcommand{\sg}[1]{{\mathrm S}_{{#1}}}
\newcommand{\va}[2]{\left\langle {#1}\right\rangle_{{#2}}}
\begin{document}
\begin{titlepage}
$\phantom{X}$\vspace{2cm}
\begin{center}
{\Large 
Renormalized $g$-$\log (g)$ double expansion for the \\[2mm]
invariant $\phi^4$-trajectory in three dimensions \\[10mm]
}
\end{center}
\begin{center}
{\Large C. Wieczerkowski} \\[10mm]
\end{center}
\begin{center}
Institut f\"ur Theoretische Physik I,
Universit\"at M\"unster, \\
Wilhelm-Klemm-Stra\ss e 9, D-48149 M\"unster, \\
wieczer@@uni-muenster.de \\[2mm]
\end{center}
\vspace{-10cm}
\hfill MS-TP1-97-02
\vspace{11cm}
\begin{abstract}\noindent
We study the invariant unstable manifold of the trivial 
renormalization group fixed point tangent to the 
$\phi^{4}$-vertex in three dimensions. We parametrize it
by a running $\phi^{4}$-coupling with linear step 
$\beta$-function. It is shown to have a renormalized
double expansion in the running coupling and its logarithm.
\end{abstract}
\end{titlepage}
\section{Introduction}

We study massless $\phi^{4}$-theory on three dimensional 
Euclidean space-time by means of Wilson's renormalization 
group, see \cite{W70,W71,WF72,WK74}. Its renormalization 
is analyzed as a dynamical system on a space of effective 
actions. The renormalization group acts on this space as 
a semi-group of scale transformations. The renormalized 
theory comes as a special one-dimensional curve of effective 
actions. It is one-dimensional as a renormalization group 
orbit. We parametrize it by a running $\phi^{4}$-coupling.  

We consider a discrete renormalization group of 
transformations $\rg{L}$, which scale by $L>1$, built from 
a momentum space decomposition of a massless Gaussian 
random field by means of an exponential regulator. See 
Gallavotti's review \cite{G85} for its application to 
renormalized perturbation theory; and Brydges' review
\cite{B92} for its use in constructive field theory. 
The renormalization group acts here on a space of potentials 
$V(\Phi)$\footnote{
$V(\Phi)$ corresponds to a perturbation ${\mathrm d}
\mu_{{\mathrm C}_{\infty}}(\Phi)\exp (-V(\Phi))$ of a 
Gaussian measure ${\mathrm d}\mu_{{\mathrm C}_{\infty}}(\Phi)$, 
with mean zero and covariance ${\mathrm C}_{L}=\int_{1}^{L^{2}}
{\mathrm d}\alpha\exp (\alpha\bigtriangleup)$, in the limit 
$L\nearrow\infty$; where $\bigtriangleup$ denotes the Laplace 
operator.} with suitable properties. 

The free massless field corresponds to the trivial fixed point 
$V_{\star}(\Phi)=0$. The linearized renormalization group at 
this trivial fixed point is diagonalized by normal ordering. 
The normal ordered local $\phi^{4}$-vertex is a particular 
eigenvector. Its scaling dimension is one in three dimensions. 
It defines an unstable perturbation of the trivial fixed point, 
which extends tangentially to an invariant curve in the unstable 
manifold of the trivial fixed point. This curve is the 
renormalized $\phi^{4}_{3}$-trajectory. We parametrize it by 
$g$ such that
\begin{equation}
\po{\Phi}{g}=
\frac{g}{4!}\int{\mathrm d}^{3}x\;:\Phi (x)^{4}:+
O(g^{2}).
\label{in1}
\end{equation}
$\log (g)$ corrections are neglected for the moment.
The program of this paper is a study of this curve as a 
function of $g$ for small couplings,  $g <<1$, beyond the 
linear approximation.

Our main dynamical principle is invariance under the 
renormalization group. As in \cite{W97}, we look for 
a pair, given by the potential $\po{\Phi}{g}$ and a step 
$\beta$-function $\bigtriangleup\beta_{L}(g)$, such that
\begin{equation}
\rg{L}(V)(\Phi\vert g)=V(\Phi\vert\bigtriangleup
\beta_{L}(g)). 
\label{in2}
\end{equation}
To first order in $g$, $\rg{L}$ is to be linearized. 
Consequently, $\bigtriangleup\beta_{L}(g)=Lg+O(g^{2})$
on the renormalized $\phi^{4}_{3}$-trajectory.\footnote{
The renormalization flow is therefore asymptotically free in
the backward direction, yielding ultraviolet asymptotic
freedom.} 
The meaning of $g$ is twofold. It is both a coordinate of 
the renormalized $\phi^{4}_{3}$-trajectory and a running 
coupling. We select it by the condition that the step  
$\beta$-function\footnote{
Infrared properties are then encoded in the strong 
coupling limit $g\nearrow\infty$.
} 
be exactly linear, 
\begin{equation}
\bigtriangleup\beta_{L}(g)=Lg.
\label{in3}
\end{equation}
The linear step $\beta$-function defines a normal form 
of the renormalization group at the trivial fixed point;
in the sense of Ecalle's theory of resurgent functions. 
Its use in renormalization theory was proposed by 
Eckmann and Wittwer, \cite{EW84}. In this representation, 
we look for $V(\Phi\vert g)$ as a fixed point of the 
composed transformation $\rg{L}\times\bigtriangleup
\beta_{L^{-1}}^{\star}$.

We will then show that the local $\phi^{2}$-vertex has a 
resonance in the order $g^{2}$, analogous to the 
hierarchical approximation \cite{RW96}; there exists no 
solution $\po{\Phi}{g}$ of \rf{in3} as a formal power 
series in $g$. As in the hierarchical approximation, 
we will then resolve the resonance by a local 
$\phi^{2}$-vertex of the type $g^{2}\;\log (g)$. This 
leads us to investigate a double expansion in $g$ and 
($g^{2}\times$) $\log(g)$. We will show that there 
exists a one parameter set of fixed points 
$\po{\Phi}{g}$ in formal double expansion. The present 
paper extends the previous study \cite{W97} of the 
renormalized $\phi^{4}_{4}$-trajectory. The closest 
relative to our method in the literature is the 
$\beta$-functional method in the tree expansion of 
Gallavotti and Nicol\`o. See \cite{BG95} for an 
introduction to this technology. However, we do not begin 
from a regularized theory defined by a bare action. 
This scheme is here replaced by a construction of an 
effective theory on all length scales in units of the 
coupling.   

The ultraviolet limit of $\phi^{4}_{3}$-theory has been 
rigorously studied in constructive field theory. See 
Glimm and Jaffe \cite{GJ73,GJ87}, Feldman and Osterwalder
\cite{FO76}, Magnen and Seneor \cite{MS77}, 
Balaban \cite{B83}, Benfatto et al. \cite{BCGNOPS80}, 
Brydges et al. \cite{BDH93}, and references therein. 
In view of this knowledge, our contribution is modest,
clarifying $\log (g)$ corrections from the dynamical 
systems point of view. Double expansions in $g$ and 
$\log (g)$ appeared also in Symanzik's work on 
non-renormalizable $\phi^{4}$-theory in $4+\epsilon$
dimensions \cite{S75}. 

\section{Renormalization group}

We study perturbations of a scalar Gaussian massless random 
field by means of a renormalization group with Gaussian 
regulator. See \cite{G85,FFS92,BG95,R91,GK84} for background
material. We will use the setup as in \cite{W97} with the
difference that the dimension of Euclidean space-time will 
be three rather than four.  

\subsection{Renormalization group transformation}

We consider the renormalization group of transformations $\rg{L}$, 
given by 
\begin{equation}
\rg{L}(V)(\Phi)=
-\log\left(\ga{{\mathrm C}_{L}}{\phi}
\exp\left(-V\left(\sg{L}(\Phi)+\phi\right)\right)
\right)-{\mathrm{const.}},
\label{rg1}
\end{equation}
where $\phi$ is a Gaussian random field with mean zero and 
covariance
\begin{equation}
C_{L}(x,y)=\ip{p}\;\ex{{\mathrm i}p(x-y)}\;
\frac{\widehat{\chi}(p)-\widehat{\chi}(Lp)}{p^{2}},
\quad\widehat{\chi}(p)=\ex{-p^{2}},
\label{rg2}
\end{equation}
and where $\sg{L}(\Phi)(x)=L^{-1/2}\;\Phi (L^{-1}x)$. A few 
remarks on the covariance \rf{rg2} are collected in the 
appendix \ref{co}. The scale $L$ will be kept fixed at some 
value $L>1$, say $L=2$. We thus consider a discrete version 
of the renormalization group. 

The covariance ${\mathrm C}_{L}$ and the dilatation $\sg{L}$ 
are related to an $L$-independent covariance 
${\mathrm C}_{\infty}$, given by
\begin{equation}
C_{\infty}(x,y)=\ip{p}\;\ex{{\mathrm i}p(x-y)}\;
\frac{\widehat{\chi}(p)}{p^{2}},
\label{rg3}
\end{equation}
through ${\mathrm C}_{L}={\mathrm C}_{\infty}-
\sg{L}{\mathrm C}_{\infty}\sg{L}^{{\mathrm T}}$. It will 
serve as normal ordering covariance. The covariance \rf{rg2}
has a unit ultraviolet cutoff and an infrared cutoff $L^{-1}$. 
The covariance \rf{rg3} has a unit ultraviolet cutoff but no
infrared cutoff. 

\subsection{Space of potentials}

We consider potentials $V(\phi)$, which are given by power 
series
\begin{equation}
V(\Phi)=
\sum_{n=1}^{\infty}\;
\frac{1}{(2n)!}\;
\ix{x_{1}}\cdots\ix{x_{2n}}\;
:\Phi (x_{1})\cdots\Phi (x_{2n}):
\;V_{2n}(x_{1},\ldots,x_{2n})
\label{rg4}
\end{equation}
in $\Phi$. The real space kernels $V_{2n}(x_{1},\ldots,x_{2n})$            
will be required to be symmetric and Euclidean invariant 
distributions, which are given by Fourier integrals
\begin{eqnarray}
V_{2n}(x_{1},\ldots,x_{2n})&=&
\ip{p_{1}}\cdots\ip{p_{2n}}\;
(2\pi)^{3}\delta (p_{1}+\cdots+p_{2n})\;
\nonumber\\&\phantom{=}&
\ex{{\mathrm i}(p_{1}x_{1}+\cdots+p_{2n}x_{2n})}
\;\widehat{V}_{2n}(p_{1},\ldots,p_{2n})
\label{rg5}
\end{eqnarray}
of smooth momentum space kernels $\widehat{V}_{2n}(p_{1},\ldots,
p_{2n})$. Smoothness will be required in particular at the 
point $p_{1}=\cdots =p_{2n}=0$. 

The matter of the convergence of \rf{rg4} as a power series in 
$\Phi$ will be left aside. We will restrict our attention 
to polynomial approximations of finite orders, leaving aside 
the large field problem. Potentials will always be normalized 
such that $V(0)=0$. The transformation \rf{rg1} is understood 
to be supplemented by a subtraction of a normalization constant, 
proportional to the volume.

\subsection{Localization operators}

We define linear localization operators $\lo{2n}$ by
\begin{equation}
\lo{2n}(V)(\Phi)=\lambda_{2n}(V)\;\ob{2n}{\Phi},\quad
\lambda_{2n}(V)=\widehat{V}_{2n}(0,\ldots,0), 
\label{rg8}
\end{equation}
where $\ob{2n}{\Phi}$ is the local $\Phi^{2n}$-vertex
\begin{equation}
\ob{2n}{\Phi}=\frac{1}{(2n)!}\ix{x}:\Phi (x)^{2n}:.
\label{rg9}
\end{equation}
This completes our setup.

\section{$\phi^{4}$-trajectory}

The $\phi^{4}$-trajectory is a renormalization invariant 
curve in the space of potentials, which emerges from the 
trivial fixed point tangent to the $\phi^{4}$-vertex. The 
$\phi^{4}$-vertex is an unstable perturbation in three 
dimensions, whereas it is a marginally stable perturbation
in four dimensions. This difference is encoded in the 
respective step $\beta$-functions. In three dimensions, the 
power counting of vertices becomes order dependent, since 
the coupling has a scaling dimension different from zero. 
Peculiar to three dimensions is a second order mass resonance, 
whose resolution requires logarithmic couplings. See 
\cite{RW96} for a treatment of its hierarchical approximation.

\subsection{Renormalization invariance}

We investigate the following renormalization problem. We 
look for a pair, consisting of a potential $\po{\Phi}{g}$
and a differential $\beta$-function $\beta (g)$, such that
\begin{equation} 
\rg{L}(V)(\Phi\vert g)=V(\Phi\vert \db{L}(g)),
\label{rt1}
\end{equation}
where $\db{L}(g)$ is the step $\beta$-function, defined 
as the solution $\db{L}(g)=g(L)$ of the ordinary 
differential equation
\begin{equation}
L\;g^{\prime}(L)=\beta (g(L)),\quad g(1)=g.
\label{rt2}
\end{equation}
A few remarks on the $\beta$-function are appended in 
\ref{be}. The $\phi^{4}$-trajectory is a particular pair, 
which is distinguished by the first order condition
\begin{equation}
\po{\Phi}{g}=V^{(1)}(\Phi)\;g+O(g^{2}),\quad
V^{(1)}(\Phi)=\ob{4}{\Phi}.
\label{rt3}
\end{equation} 
We postpone the matter of $\log (g)$ corrections in 
$\po{\Phi}{g}$ and $\beta (g)$ for the moment. 

\subsection{Order $g$}

To first order in $g$, \rf{rt1} becomes an eigenvector
problem
\begin{equation}
\va{V^{(1)}}{{\mathrm C}_{L},\sg{L}\Phi}
-L^{\beta^{(1)}}\; V^{(1)}(\Phi)=0
\label{rt4}
\end{equation}
for the linearized renormalization group
\begin{equation}
\va{{\mathcal O}}{{\mathrm C}_{L},\sg{L}\Phi}=
\ga{{\mathrm C}_{L}}{\phi}\;{\mathcal O}(\sg{L}\Phi+
\phi)
\label{rt5}
\end{equation}
with the eigenvalue related to
\begin{equation}
\beta (g)=\beta^{(1)}\;g+O(g^{2}),\quad
\db{L}(g)=L^{\beta^{(1)}}\;g+O(g^{2}).
\label{rt6}
\end{equation}
The normal ordered $\phi^{4}$-vertex $\ob{4}{\Phi}$
is an eigenvector of the linearized renormalization 
group \rf{rt5}. Its eigenvalue is $\beta^{(1)}=1$ in
three dimensions.

\subsection{$\beta$-function}

To higher orders in $g$, our renormalization problem
has to be supplemented either by a condition on  
$\po{\Phi}{g}$, by a condition on $\beta (g)$, or 
by a condition on both to define $g$. A canonical 
condition on $\po{\Phi}{g}$ is 
\begin{equation}
\lo{4}(V)(\Phi\vert g)=\ob{4}{\Phi}\;g,\quad
\widehat{V}_{4}(0,0,0,0\vert g)=g.
\label{rt7}
\end{equation}
It defines $g$ to be the local $\phi^{4}$-coupling.
With this choice, the $\beta$-function has to be 
computed from \rf{rt1}. A canonical condition on 
$\beta (g)$ is 
\begin{equation}
\beta (g)=\beta^{(1)}\;g,\quad
\db{L}(g)=L^{\beta^{(1)}}\;g, 
\label{rt8}
\end{equation}
provided that $\beta^{(1)}$ is not zero. 
With this choice, 
the local $\phi^{4}$-coupling, we name it
$\lambda_{4}(g)$, has to be computed
from \rf{rt1}. Mixed conditions will not be 
considered here. The linear $\beta$-function has 
technical advantages in the below double 
perturbation theory. It can be viewed as a 
normal coordinate in the vicinity of the trivial
fixed point. Therefore, we choose to work with 
the second condition \rf{rt8}. Some remarks on 
this issue have been added in the appendix \ref{be}. 
Our renormalization problem has now turned into
a fixed point problem. We seek to compute 
$\po{\Phi}{g}$ as a fixed point of $\rg{L}\times 
\db{L^{-1}}^{\star}$, for all $L>1$. 

\subsection{Order $g^{2}$ mass resonance}

There exists no second order potential $V^{(2)}(\Phi)$
such that the fixed point $V(\Phi\vert g)$ is of the 
form 
\begin{equation}
\po{\Phi}{g}=\ob{4}{\Phi}\;g+
V^{(2)}(\Phi)\;\frac{g^{2}}{2}+O(g^{3}).
\label{rt9}
\end{equation}
The reason is the following. Suppose that there 
existed a fixed point \rf{rt9}. To second order in 
$g$, \rf{rt1} would require that
\begin{equation}
L^{-2}\;\va{V^{(2)}}{{\mathrm C}_{L},\sg{L}\Phi}-
V^{(2)}(\Phi)=L^{-2}
\va{V^{(1)};V^{(1)}}{{\mathrm C}_{L},\sg{L}
\Phi}^{{\mathrm T}},
\label{rt10}
\end{equation}
where the right hand side is a truncated 
expectation value. Eq. \rf{rt10} implies that
\begin{equation}
L^{-2}\;\lo{2}
\va{V^{(1)};V^{(1)}}{{\mathrm C}_{L},\sg{L}
\Phi}^{{\mathrm T}}=0,
\label{rt11}
\end{equation}
because the local $\phi^{2}$-couplings on 
the left hand side of \rf{rt10} cancel. 
This proves to be false. See \ref{or}. We have a 
contradiction.

The problem is that a local $\phi^{2}$-vertex
is generated to second order, while the 
eigenvalue $L^{2}$ of $\ob{2}{\Phi}$ is the
square of the eigenvalue $L$ of $\ob{4}{\Phi}$.
This situation was called a mass resonance.  
See \cite{RW96} for a list of various 
resonances in the hierarchical approximation, 
and their resolutions. Resonances have to be 
resolved by logarithmic couplings. 

\subsection{Order $g^{2}$}

Concerning all other vertices than the local 
$\phi^{2}$-vertex, eq. \rf{rt10} has a unique
regular solution. It poses a linear problem for 
$\lo{2}^{\perp}V^{(2)}(\Phi)$, given $V^{(1)}(\Phi)$, 
where $\lo{2}^{\perp}=1-\lo{2}$. Restricted to 
$\lo{2}^{\perp}$, eq. \rf{rt10} is equivalent to a 
set of difference equations for the momentum space 
kernels \rf{rg5}. We introduce the abbreviation
\begin{equation}
\kg{L}(V)^{(2)}(\Phi)=
L^{-2}\;\va{V^{(1)};V^{(1)}}{{\mathrm C}_{L},\sg{L}
\Phi}^{{\mathrm T}},
\label{rt12}
\end{equation}
to obtain a system of equations 
\begin{equation}
L^{1-n}\;\widehat{V}_{2n}^{(2)}(L^{-1}\;p_{1},
\ldots,L^{-1}\;p_{2n})-\widehat{V}_{2n}^{(2)}
(p_{1},\ldots,p_{2n})=
\widehat{\kg{L}(V)}_{2n}^{(2)}(p_{1},\ldots,
p_{2n}).
\label{rt13}
\end{equation}
The connected normal ordered contraction of two 
normal ordered local $\phi^{4}$-vertices is a 
polynomial of degree six in $\Phi$. The right hand 
side of \rf{rt13} is therefore zero for $n>3$. 
For $n\geq 2$, eq. \rf{rt13} is solved by
\begin{equation}
\widehat{V}_{2n}^{(2)}(p_{1},\ldots,p_{2n})=
-\sum_{\alpha=0}^{\infty}\;L^{\alpha\;(1-n)}\;
\widehat{\kg{L}(V)}_{2n}^{(2)}(L^{-\alpha}\;p_{1},
\ldots,L^{-\alpha}\;p_{2n}).
\label{rt14}
\end{equation}
This solution is regular at $p_{1}=\cdots =p_{2n}=0$.
It is the only solution with this property. 
More details on this can be found in \cite{W97}.
The non-local $\lo{2}^{\perp}$-part of the 
$\phi^{2}$-vertex is given by
\begin{equation}
\widehat{\lo{2}^{\perp}(V)}_{2}^{(2)}(p_{1},p_{2})=
-\sum_{\alpha=0}^{\infty}\;
\widehat{\lo{2}^{\perp}\kg{L}(V)}_{2}^{(2)}
(L^{-\alpha}\;p_{1},L^{-\alpha}\;p_{2}).
\label{rt15}
\end{equation} 
This sum converges, despite the absence of a power 
counting factor, due to the subtraction of the local 
$\phi^{2}$-coupling; there exists a radius $r>0$ and 
a constant $C>0$ such that
\begin{equation}
\left\vert\widehat{\lo{2}^{\perp}\kg{L}(V)}^{(2)}_{2}
(p,-p)\right\vert\leq C\;p^{2}
\label{rt16.1}
\end{equation} 
for all $p^{2}\leq r^{2}$. Except for the local 
$\phi^{2}$-vertex $\lo{2}(V)^{(2)}(\Phi)$, we have
a unique regular solution of \rf{rt10}. We rename it
$\lo{2}^{\perp}(V)^{(2,0)}(\Phi)$.

There remains the issue of large momentum bounds. 
In three dimensions we have better large momentum 
bounds than in four dimensions, see \cite{W97}
concerning the latter. Indeed, since there are no 
subtractions in $\lo{2}^{\perp}$, we find that 
the right hand side of \rf{rt13}, given by 
convergent convolution integrals, is a bounded 
function of the momenta. Consequently, the solutions
\rf{rt14} obey
\begin{equation}
\|\widehat{V}_{2n}^{(2)}\|_{\infty}<\infty,\quad
n\geq 2.
\label{rt16.2}
\end{equation}
The non-local $\lo{2}^{\infty}$-part \rf{rt15}
is the only vertex in this theory which is not
a bounded function of momentum. Its accurate 
large momentum behavior is logarithmic. This bound
is explained in the appendix \ref{lm}. 
At higher orders, all vertices will be bounded. 
The reason is that the logarithmic growth of the 
order two $\phi^{2}$-vertex is more than compensated 
by the exponential decay of the propagators at 
large momentum.

\subsection{$g^{2}\;\log (g)$ mass term}

Armed with this insight, we then consider the following 
modified second order, 
\begin{eqnarray}
\po{\Phi}{g}&=&\ob{4}{\Phi}g+
\lo{2}^{\perp}(V)^{(2,0)}(\Phi)\;\frac{g^{2}}{2}+
\ob{2}{\Phi}\left\{\lambda_{2}^{(2,0)}+
\lambda^{(2,1)}\log (g)\right\}\;\frac{g^{2}}{2}
\nonumber\\&\phantom{=}&
+O(g^{3},g^{3}\log (g)),
\label{rt17}
\end{eqnarray}
where we have added a local $\phi^{2}$-coupling
\begin{equation}
\lambda_{2}(g)=
\lambda^{(2,0)}\;\frac{g^{2}}{2}+
\lambda^{(2,1)}\;\frac{g^{2}}{2}\log (g)+
O(g^{3},g^{3}\log (g)).
\label{rt18}
\end{equation}
The point with it is that
\begin{equation}
L^{-2}\;\lambda_{2}(Lg)=
\left\{\lambda_{2}^{(2,0)}+\log (L)\;\lambda_{2}^{(2,1)}
\right\}\frac{g^{2}}{2}+\lambda_{2}^{(2,1)}\;
\frac{g^{2}}{2}\;\log (g)+O(g^{3},g^{3}\log (g)).
\label{rt19}
\end{equation}
We then consider separately the order $g^{2}$-term and 
the order $g^{2} \log (g)$-term in \rf{rt1}. They 
are given by
\begin{equation}
L^{-2}\;\va{V^{(2,0)}}{{\mathrm C}_{L},\sg{L}\Phi}-
V^{(2,0)}(\Phi)-
\lambda_{2}^{(2,1)}\;\log (L)\;\ob{2}{\Phi}=L^{-2}
\va{V^{(1,0)};V^{(1,0)}}{{\mathrm C}_{L},\sg{L}
\Phi}^{{\mathrm T}},
\label{rt20}
\end{equation}
together with
\begin{equation}
L^{-2}\;\va{V^{(2,1)}}{{\mathrm C}_{L},\sg{L}\Phi}-
V^{(2,1)}(\Phi)=0.
\label{rt21}
\end{equation}
The flow of the logarithmic local $\phi^{2}$-vertex
cures the resonance. Apply $\lo{2}$ to \rf{rt20} to
obtain
\begin{equation}
\lambda_{2}^{(2,1)}\;\log(L)\;\ob{2}{\Phi}=
L^{-2}\;\va{V^{(1,0)};V^{(1,0)}}{{\mathrm C}_{L},\sg{L}
\Phi}^{{\mathrm T}}.
\label{rt22}
\end{equation}
This equation determines the invariant coupling 
parameter $\lambda_{2}^{(2,1)}$. The non-logarithmic
second order parameter $\lambda_{2}^{(2,0)}$ is 
not determined by \rf{rt20} and \rf{rt21}. It is a 
free parameter of this theory. All higher order 
vertices, both the logarithmic and the non-logarithmic
ones, have negative power counting. They are thus 
irrelevant, and therefore uniquely determined. Thus 
$\lambda_{2}^{(2,0)}$ is the only free parameter of 
this renormalization problem. The curve with a minimal
set of vertices is $\lambda_{2}^{(2,0)}=0$. 

\section{Perturbation theory}

We added a $g^{2}\;\log (g)$ correction to the local 
$\phi^{2}$-vertex. It suggests a double expansion in 
$g$ and ($g^{2}\times$) $\log (g)$ for the 
$\phi^{4}$-trajectory. The program of this section 
is this double expansion. It will be formulated in 
terms of a recursion relation. When the recursion
relation is iterated, $\log (g)$-corrections spread
to all higher vertices.

\subsection{$g$-$\log (g)$ double expansion}

We consider potentials of the general form
\begin{equation}
\po{\Phi}{g}=\sum_{r=1}^{\infty}\;
\sum_{a=1}^{\left[\frac{r}{2}\right]}\;
V^{(r,a)}(\Phi)\;\frac{g^{r}}{r!}\;
\frac{\log (g)^{a}}{a!},
\label{pt1}
\end{equation}
where $\left[\frac{r}{2}\right]$ denotes the integer
part of $\frac{r}{2}$, and where $V^{(r,a)}(\Phi)$
is a polynomial 
\begin{equation}
V^{(r,a)}(\Phi)=
\sum_{n=1}^{r-2a+1}\;
\frac{1}{(2n)!}\;\ix{x_{1}}\cdots\ix{x_{2n}}\;
:\Phi (x_{1})\cdots\Phi (x_{2n}):
\;V^{(r,a)}_{2n}(x_{1},\ldots,x_{2n})
\label{pt2}
\end{equation}
in $\Phi$. The order $r-2a+1$ is particular to 
$\phi^{4}$-theory. It equals half the maximal 
number of external legs of all connected Feynman 
graphs, which can be built from $r-2a$ local 
$\phi^{4}$-vertices and $a$ local 
$\phi^{2}$-vertices. We anticipate that this form
of potentials iterates through the renormalization
group. The real space vertices $V^{(r,a)}_{2n}(x_{1},
\ldots,x_{2n})$ are required to be given by Fourier
integrals of the form \rf{rg5}.

\subsection{Flow of $g$}

The step $\beta$-function is $\db{L}(g)=L\;g$ in 
the present representation. The right hand side of 
\rf{rt1} then becomes  
\begin{equation}
\po{\Phi}{Lg}=\sum_{r=1}^{\infty}\;
\sum_{a=1}^{\left[\frac{r}{2}\right]}\;
\left\{L^{r}\;\sum_{b=a}^{\left[\frac{r}{2}\right]}\;
\frac{\log (L)^{b-a}}{(b-a)!}\;V^{(r,b)}(\Phi)
\right\}\;\frac{g^{r}}{r!}\;
\frac{\log (g)^{a}}{a!}.
\label{pt3}
\end{equation}               
Potentials, which are given by \rf{pt1} and \rf{pt2},
thus remain of this form under our flow of $g$.  

\subsection{Renormalization transformation}

The image of a renormalization transformation \rf{rg1},
applied to the potential \rf{pt1}, is 
\begin{equation} 
\rg{L}(V)(\Phi)=
\sum_{r=1}^{\infty}\sum_{a=1}^{\left[\frac{r}{2}\right]}
\rg{L}(V)^{(r,a)}(\Phi),
\label{pt4}
\end{equation}
where $\rg{L}(V)^{(r,a)}(\Phi)$ is again a polynomial 
of the form \rf{pt2}. The maximal order $r-2a+1$ thus
iterates through the renormalization transformation.
The coefficients in \rf{pt4} are given by sums of 
truncated expectation values
\begin{gather}
\rg{L}(V)^{(r,a)}(\Phi)=
\sum_{i=1}^{r}\;
\frac{(-1)^{i+1}}{i!}\;
\sum_{(s_{1},\ldots,s_{i})\in\se{N}^{i}}\;
\delta_{\sum_{j=1}^{i}s_{j}-r}\;
\binom{r}{s_{1}\cdots s_{i-1}}\;
\nonumber\\
\sum_{(b_{1},\ldots,b_{i})\in\se{N}^{i}}\;
\delta_{\sum_{j=1}^{i}b_{j}-a}\;
\binom{a}{b_{1}\cdots b_{i-1}}\;
\va{V^{(s_{1},b_{1})};\ldots;V^{(s_{i},b_{i})}}
{{\mathrm C}_{L},\sg{L}(\Phi)}^{\mathrm T},
\label{pt5}
\end{gather} 
with the agreement that $V^{(s,b)}(\Phi)=0$ for 
$b>\left[\frac{s}{2}\right]$ or $s=0$. 

\subsection{Scaling equations}

Equate \rf{pt3} and \rf{pt4} to obtain a linear
equation
\begin{equation}
L^{-r}\;\va{V^{(r,a)}}{{\mathrm C}_{L},\sg{L}(\Phi)}
-V^{(r,a)}(\Phi)=\kg{L}(V)^{(r,a)}(\Phi)
\label{pt6}
\end{equation}
for $V^{(r,a)}(\Phi)$, whose right hand side is 
the non-linear transformation
\begin{gather}
\kg{L}(V)^{(r,a)}(\Phi)=
\sum_{b=a+1}^{\left[\frac{r}{2}\right]}
\frac{\log (L)^{b-a}}{(b-a)!}\;
V^{(r,b)}(\Phi)-L^{-r}\;
\sum_{i=2}^{r}\frac{(-1)^{i+1}}{i!}\;
\nonumber\\
\sum_{(s_{1},\ldots,s_{i})\in\se{N}^{i}}\;
\delta_{\sum_{j=1}^{i}s_{j}-r}\;
\binom{r}{s_{1}\cdots s_{i-1}}\;
\sum_{(b_{1},\ldots,b_{i})\in\se{N}^{i}}\;
\delta_{\sum_{j=1}^{i}b_{j}-a}\;
\binom{a}{b_{1}\cdots b_{i-1}}\;
\nonumber\\
\va{V^{(s_{1},b_{1})};\ldots;V^{(s_{i},b_{i})}}
{{\mathrm C}_{L},\sg{L}(\Phi)}^{\mathrm T}.
\label{pt7}
\end{gather}
The computation of the order $(r,a)$, with 
$r\geq 3$ and $0\leq a\leq\left[\frac{r}{2}\right]$, 
requires knowledge of the lower $g$ orders $(s,b)$, 
with $1\leq s\leq r-1$ and $0\leq b\leq\left[\frac{s}{2}
\right]$, and of the higher $\log (g)$ orders $(r,b)$, 
with $a+1 \leq b\leq\left[\frac{r}{2}\right]$. 
This is achieved by double recursion, which proceeds
forward in the $g$ orders and backwards in the 
$\log (g)$ orders.
 
\subsection{Scaling vertices}

If all previous orders $(s,b)$, which needed to compute 
\rf{pt7} in the double recursion, are of the polynomial 
form \rf{pt2}, then also \rf{pt2} is of this polynomial 
form
\begin{eqnarray}
\kg{L}(V)^{(r,a)}(\Phi)&=&
\sum_{n=1}^{r-2a+1}\;
\frac{1}{(2n)!}\;\ix{x_{1}}\cdots\ix{x_{2n}}\;
:\Phi (x_{1})\cdots\Phi (x_{2n}):
\nonumber\\& &\quad
\;\kg{L}(V)^{(r,a)}_{2n}(x_{1},\ldots,x_{2n}).
\label{pt8}
\end{eqnarray}
The maximal order $r-2a+1$ thus iterates through the
renormalization group. If the real space kernels of 
all previous vertices are given by Fourier integrals 
\rf{rg5} of smooth momentum space kernels, then this
is also true for
\begin{eqnarray}
\;\kg{L}(V)^{(r,a)}_{2n}(x_{1},\ldots,x_{2n})&=&
\ip{p_{1}}\cdots\ip{p_{2n}}\;
(2\pi)^{3}\delta (p_{1}+\cdots+p_{2n})\;
\nonumber\\&\phantom{=}&\
\ex{{\mathrm i}(p_{1}x_{1}+\cdots+p_{2n}x_{2n})}
\;\widehat{\kg{L}(V)}_{2n}^{(r,a)}(p_{1},\ldots,p_{2n}).
\label{pt9}
\end{eqnarray}
There is a technical point here. One has to show a 
large momentum bound on the momentum space kernels 
to ensure the convergence of the convolutions 
involved in \rf{pt9}. An exponential bound suffices, 
since the propagators \rf{rg2} and \rf{rg3} both 
have exponential ultraviolet cutoffs. The estimates 
of \cite{W97} directly apply and yield such an 
exponential bound. But we have a better bound in 
this three dimensional case. The scaling equation 
\rf{pt6} becomes a system of difference equations
\begin{equation}
L^{3-r-n}\;\widehat{V}_{2n}^{(r,a)}(L^{-1}\;p_{1},
\ldots,L^{-1}\;p_{2n})-\widehat{V}_{2n}^{(r,a)}
(p_{1},\ldots,p_{2n})=
\widehat{\kg{L}(V)}_{2n}^{(r,a)}(p_{1},\ldots,
p_{2n}).
\label{pt10}
\end{equation}
We solve it inductively. The right hand side can be
assumed to consist of symmetric, Euclidean invariant,
and smooth momentum space kernels. The unique 
solution of \rf{pt10}, which has again these 
properties, is
\begin{equation}
\widehat{V}_{2n}^{(r,a)}(p_{1},\ldots,p_{2n})=
-\sum_{\alpha=0}^{\infty}\;L^{\alpha\;(1-n)}\;
\widehat{\kg{L}(V)}_{2n}^{(r,s)}(L^{-\alpha}\;p_{1},
\ldots,L^{-\alpha}\;p_{2n}).
\label{pt11}
\end{equation}
In other words, we have an explicit recursion relation,
from which we obtain information about all orders of
formal double perturbation expansion. This theory has
one free parameter, the local $\phi^{2}$-coupling at
order $(2,0)$. 

The scaling dimension $3-r-n$ in \rf{pt10} is negative
in all orders $r\geq 3$, which we are now concerned 
with. Differing from the renormalizable case \cite{W97}, 
no subtractions are necessary in higher orders. The 
vertices are all irrelevant. The right hand side of
\rf{pt9} is a sum of convergent convolution integrals,
where all vertices except for non-local second 
order $\phi^{2}$-vertex are bounded. Consequently, 
the right hand side of \rf{pt9} is bounded for all
triples $(r,a,n)$ with $r\geq 3$. Therefore,
\begin{equation}
\|\widehat{V}_{2n}^{(r,a)}\|_{\infty}<\infty
\label{pt12}
\end{equation} 
iterates through the recursion \rf{pt10}.

\section{Summary}

We have computed the renormalized $\phi^{4}_{3}$-trajectory 
in a double a double expansion in $g$ and $\log(g)$ with
linear step $\beta$-function for a discrete renormalization
group as a fixed point of a composed transformation, 
consisting of a renormalization group transformation and
a backward flow of the coupling. Such a double expansion in 
$g$ and $\log (g)$ can be carried out also for the continuous 
renormalization group of Wilson \cite{WK74} and Polchinski 
\cite{P84}. We have chosen a discrete version as a first
step towards a constructive analysis. The double expansion
is expected to give accurate results for a small coupling
region $g<<1$. It cannot be expected to tell much about the 
strong coupling region $g>>1$. If there is no
dynamical mass generation, the limit $g\nearrow\infty$ 
should approach the non-trivial Wilson fixed point. A
convincing analysis of this limit would be highly desirable.

\section{Appendix}

\subsection{Covariance ${\mathrm C}_{L}$}
\label{co}

The covariance $C_{L}$ has both a Gaussian ultraviolet 
cutoff and a Gaussian infrared cutoff. It has the integral 
representation
\begin{equation}
C_{L}(x,y)=\int_{1}^{L^{2}}{\mathrm d}\alpha\;
(4\pi\alpha)^{\frac{-3}{2}}\;\ex{\frac{-(x-y)^{2}}{4\alpha}}.
\label{co1}
\end{equation}
Consequently, $C_{L}(x,y)$ is real, positive, and satisfies
the bound
\begin{equation}
C_{L}(x,y)\leq 
\frac{1-L^{-1}}{4\pi^{3/2}}\ex{\frac{-(x-y)^{2}}{4L^{2}}}.
\label{co2}
\end{equation}
The Fourier transformed covariances $\widehat{C}_{L}$ and 
$\widehat{C}_{\infty}$ satisfy the bounds 
\begin{equation}
\| \widehat{C}_{L}\|_{\infty}<\infty,\quad
\| \widehat{C}_{L}\|_{1}<\infty,\quad 
\| \widehat{C}_{\infty}\|_{1}<\infty.
\label{co3}
\end{equation}
See \cite{R91} for further information on \rf{co1}.

\subsection{$\beta$-function}
\label{be}

Step $\beta$-functions $\db{L}$ are required to satisfy the
composition law
\begin{equation}
\db{L}\circ\db{L^{\prime}}=\db{LL^{\prime}}.
\label{be1}
\end{equation} 
It follows that the associated running coupling, defined by
$g(L)=\db{L}(g)$, is the solution of the ordinary differential
equation
\begin{equation}
L\;g^{\prime}(L)=\beta (g(L)),\quad g(1)=g,
\label{be2}
\end{equation}
where $\beta$ is the differential $\beta$-function
\begin{equation}
\beta (g)=\partial_{L}\db{L}(g)\biggr\vert_{L=1}.
\label{be3}
\end{equation}
Step $\beta$-functions and differential $\beta$-functions 
are in one-to-one correspondence.

Differential $\beta$-functions transform under
coordinate transformations $\overline{g}(g)$ as vector 
fields, 
\begin{equation}
\overline{\beta}(\overline{g})=
\beta (g)\;\partial_{g}\overline{g}(g).
\label{be4}
\end{equation}
Suppose that our differential $\beta$-function behaves 
for small couplings as
\begin{equation}
\beta (g)=\beta^{(1)}\;g+O(g^{2})
\label{be5}
\end{equation}
with $\beta^{(1)}>0$. This first coefficient then turns
out to be universal. The normal form of this 
differential $\beta$-function turns out to be the 
linear function
\begin{equation}
\overline{\beta}(\overline{g})=\beta^{(1)}
\;\overline{g}.
\label{be6}
\end{equation}
To see this, it suffices to consider coordinate 
transformations, which behave like
\begin{equation}
\overline{g}(g)=\overline{g}^{(1)}\;g+O(g^{2})
\label{be7}
\end{equation}
for small couplings. Eqs. \rf{be4} and \rf{be6} yield 
a differential equation for $\overline{g}(g)$. It is 
integrated to
\begin{equation}
\overline{g}(g)=\overline{g}^{(1)}\;
\exp\left(\beta^{(1)}\;\int_{0}^{g}
\frac{{\mathrm d}g^{\prime}}{\beta (g^{\prime})}
\right).
\label{be8}
\end{equation}
The coefficient $\overline{g}^{(1)}$ is undetermined. 
We can choose $\overline{g}^{(1)}=1$, for instance. 
For small couplings, the coordinate transformation 
then reads
\begin{equation}
\overline{g}(g)=g+\frac{\beta^{(2)}}{\beta^{(1)}}\;
\frac{g^{2}}{2}+O(g^{3}).
\label{be9}
\end{equation}
As a formal power series in $g$, it is computed by 
differentiating \rf{be8}. It follows that there 
always exists such a coordinate transformation in 
the sense of a formal power series. Beyond 
perturbation theory there remains the question of 
the convergence of \rf{be8}. With an additional  
bound on the $O(g^{2})$ corrections in \rf{be4}, 
its existence can be proved for small couplings. 

\subsection{Order $g^{2}$}
\label{or}

To order $g$, we selected the potential 
\begin{equation}
\po{\Phi}{g}=V^{(1,0)}(\Phi)\;g
+O(g^{2},g^{2}\log (g)),\quad
V^{(1,0)}(\Phi)=\ob{4}{\Phi}.
\label{or1}
\end{equation}
To order $g^{2}$, we are then led to compute 
the truncated expectation value
\begin{gather}
\va{{\mathcal O}_{4};{\mathcal O}_{4}}{{\mathrm C}_{L},
\sg{L}\Phi}^{{\mathrm T}}=\ix{x}\ix{y}\biggl\{
\frac{1}{2}:\Phi(x)\Phi(y):\;L^{5}\ob{2}{L(x-y)}+
\nonumber\\
\frac{1}{4!}:\Phi(x)^{2}\Phi(y)^{2}:\;L^{4}\ob{4}{L(x-y)}+
\frac{1}{6!}:\Phi(x)^{3}\Phi(y)^{3}:\;L^{3}\ob{6}{L(x-y)}
\biggr\},
\label{or2}
\end{gather}
with real space kernels
\begin{eqnarray}
\ob{2}{x-y}&=&\frac{1}{3}C_{L}(x-y)^{3}+
C_{L}(x-y)^{2}\;L^{-1}C_{\infty}(L^{-1}(x-y))+
\nonumber\\&\phantom{=}&
C_{L}(x-y)\;L^{-2}C_{\infty}(L^{-1}(x-y))^{2},
\label{or3}\\
\ob{4}{x-y}&=&3\;C_{L}(x-y)^{2}+
6\;C_{L}(x-y)\;L^{-1}C_{\infty}(L^{-1}(x-y)),
\label{or4}\\
\ob{6}{x-y}&=&20\;C_{L}(x-y).
\label{or5}
\end{eqnarray}
Define $C^{L}_{\infty}(x-y)=L^{-1}C_{\infty}(L^{-1}(x-y))$.
The momentum space kernels are then given by convolutions
\begin{eqnarray}
\widehat{{\mathcal O}}_{2}(p)&=&
\frac{1}{3}\widehat{C}_{L}\star\widehat{C}_{L}\star
\widehat{C}_{L}(p)+
\widehat{C}_{L}\star\widehat{C}_{L}\star
\widehat{C}_{\infty}^{L}(p)+
\widehat{C}_{L}\star\widehat{C}_{\infty}^{L}\star
\widehat{C}_{\infty}^{L}(p), 
\label{or6}\\
\widehat{{\mathcal O}}_{4}(p)&=&
3\;\widehat{C}_{L}\star\widehat{C}_{L}(p)+
6\;\widehat{C}_{L}\star\widehat{C}_{\infty}^{L}(p),
\label{or7}\\
\widehat{{\mathcal O}}_{6}(p)&=&
20\;\widehat{C}_{L}(p).
\label{or8}
\end{eqnarray}
It follows in particular that $\widehat{{\mathcal O}}_{2}(0)$
is different from zero. The $g^{2}\log (g)$ local 
$\phi^{2}$-vertex is thus indeed generated.

\subsection{Logarithmic bound}
\label{lm}

Let $G(p)$, $p\in\se{R}$ for simplicity, be a 
given function such that
\begin{equation}
G(0)=0,\quad G\in{\mathcal C}^{\infty}(\se{R}),
\quad \| G\|_{\infty}<\infty.
\label{lm1}
\end{equation}
Let $F(p)$ be the solution
\begin{equation}
F(p)=-\sum_{\alpha=0}^{\infty}\;G(L^{-\alpha}p)
\label{lm2}
\end{equation}
of the difference equation 
\begin{equation}
F(L^{-1}p)-F(p)=G(p),
\label{lm3}
\end{equation}
for a given value of $L$ such that $L>1$.

The series \rf{lm2} converges uniformly on 
compact subsets of $\se{R}$. It defines a 
function $F$ with the properties
\begin{equation}
F(0)=0,\quad F\in{\mathcal C}^{\infty}(R),
\label{lm4}
\end{equation}
and 
\begin{equation}
\vert F(p)\vert\leq
\begin{cases}
A\;\vert p\vert, & \vert p\vert\leq 1, \\
B+C\;\log\vert p\vert, & \vert p\vert > 1,
\end{cases}
\label{lm5}
\end{equation}
with 
\begin{equation}
A=\frac{1}{1-L^{-1}}\;
\sup_{\vert q\vert\leq 1} \vert G^{\prime}(q)\vert,\quad
B=A+\| G\|_{\infty},\quad
C=\frac{\| G\|_{\infty}}{\log (L)}.
\label{lm6}
\end{equation}
The bound \rf{lm5} follows inductively from \rf{lm3}. 
For $\vert p\vert\leq 1$, we have that 
\begin{equation}
\vert G(p)\vert\leq \sup_{\vert q\vert\leq 1}
\vert G^{\prime}(q)\vert\;\vert p\vert.
\label{lm7}
\end{equation}
The first case in \rf{lm5} follows. For $1<\vert p\vert
\leq L$, then \rf{lm3} implies that
\begin{equation}
\vert F(p)\vert\leq
A\;L^{-1}\vert p\vert+\| G\|_{\infty}\leq
B+C\log\vert p\vert.
\label{lm8}
\end{equation}
We then proceed from $L^{n-1}<\vert p\vert\leq L^{n}$ to 
$L^{n}<\vert p\vert\leq L^{n+1}$ using \rf{lm3} to 
obtain 
\begin{equation}
\vert F(p)\vert\leq
B+C\log \vert p\vert-C\log (L)+\| G\|_{\infty}
=B+C\log\vert p\vert,
\label{lm9}
\end{equation}
proving the second case in \rf{lm5}.\\[2mm]
{\bf Acknowledgements.} I would like to thank G.\,M\"unster,
A.\,Pordt, J.\,Rolf, and P.\,Wittwer for helpful discussions.

\end{document}